\documentclass{article}
\usepackage{arxiv}
\usepackage[utf8]{inputenc} 
\usepackage[T1]{fontenc}    
\usepackage{url}            
\usepackage{booktabs}       
\usepackage{amsfonts}       
\usepackage{nicefrac}       
\usepackage{microtype}      
\usepackage{url}
\usepackage{array}
\usepackage{amsmath,amssymb,amsfonts}
\usepackage{algorithmic}
\usepackage{physics}
\usepackage{graphicx}
\usepackage{textcomp}
\usepackage{stackrel}
\usepackage{xcolor,colortbl}
\usepackage{comment}
\usepackage{xcolor}
\usepackage{booktabs}
\usepackage{graphicx}
\usepackage{doi}
\usepackage{setspace}
\usepackage{mathtools}
\usepackage{wasysym}
\usepackage{latexsym}

\title{Experimental demonstration  of  high-entropy time of arrival based optical QRNG qualifying stringent statistical tests}
\date{} 					
        
\author{ {Anindita Banerjee \thanks{aninditabanerjee.physics@gmail.com}} 
		\And
	{Anuj Sethia} 
	\And
	{Vijayalaxmi Mogiligidda} 
	\And
	{Rajesh Kumar Krishnan  } 
		\And
	{Meiyappan AR  } 
		\And
	{Sairam Rajamani  } 
		\And
	{Vivek Shenoy  }
	\And {            }
	\And {            }
	\And {            }
		\centering{QuNu Labs Pvt. Ltd.,
              M.G. Road, Bangalore, India}
              }

\begin{document}
\maketitle

\begin{abstract}
We report a demonstration of a high-entropy optical quantum random number generator (QRNG) based on photon  arrival time. We have implemented the scheme with  high-speed and high-precision electronics with a  time resolution of 1 ps generating  115 Mbps raw data. The random bit generation efficiency  is 8 bits per detection. The experimental data is quite consistent with theoretical estimation showing minimum bias. We apply a real-time information-theoretic randomness extractor   to generate a final data rate of 109 Mbps.  The randomness is rigorously evaluated against  well-known statistical test suites of NIST, ENT, Diehard, TU-01 and Dieharder.
\end{abstract}


\section{Introduction}

The increasing demand for high entropy random numbers for stronger computational capabilities in  science, engineering, and data protection is of global importance. The state-of-the-art random number generators utilize a perfect source of entropy based on inherent randomness of quantum mechanics \cite{HerreroCollantes}. Traditionally, complex mathematical algorithms generate random numbers, known as pseudo-random number generators (PRNGs). These algorithms consume a short random seed and produce a long sequence of numbers that appear random. Its security is based on the high entropy and secrecy of the seed. True random number generators (TRNG) generate numbers by collecting uncorrelated data from physical processes. QRNGs are a particular class of TRNG, producing random numbers from a quantum process. The inherent randomness of the quantum process is considered an unconditional and sustainable entropic source. 

In 1956, Ishida \textit{et al.} \cite{ishida1956} provided a stepping stone by utilizing the radioactivity of Cobalt-60 as an accessible source of quantum randomness. Thereafter, several proposals of QRNG came into existence \cite{HerreroCollantes}. The most promising QRNGs in terms of throughput and flexibility relies on quantum optics. The randomness extraction in  optical QRNG based on photon arrival times  and the QRNG based on radioactive decay are based on random arrival occurrences of events \cite{HerreroCollantes}. 

Recently, the arrival time information of the photon at the detector is used to generate high-speed random numbers \cite{Wayne2009, Wahl2011, Li2013, Wayne2010, Ma2005}. Compared to the path-based QRNG scheme, whose generation rate is bound by the detector count rate, the speed of this photon arrival time-based QRNG scheme can reach n-fold higher, where $N$ depends on the time resolution of the measurement. The $N$-fold increase in information extraction from a quantum state enables high-speed random number generator. In these works, photons emitted from a continuous laser diode are measured by a single-photon detector and the time intervals between successive detection events are recorded as the raw data. However, the arrival time of photon follows a Poisson distribution hence, to obtain uniform distribution we require  post-processing which   reduces the length of final random numbers. 

Researchers have investigated different approaches  to eliminate this bias and create a uniform distribution. Nie \textit{et al.} \cite{Nie2014} have generated quantum random numbers relative to an external reference clock.  Instead of measuring the time difference of consecutive photons, the time difference between photon detection and an external time reference is measured as the raw data. The method generates approximate uniform random digits based on the precision of time resolution. This quantity is shown to follow a uniform distribution. 

In this work, we present a self-contained QRNG device with a raw data throughput of 115 Mbps. This article contains seven sections. In section \ref{Optical source}, we describe the optical source and understand the photon statistics. In section \ref{Source of randomness}, a mathematical model of randomness source and probability model is described. The experimental setup  is explained in section \ref{system_architechture} and the implementation of scheme is presented in section \ref{implementation}. Section \ref{bias} discusses  the possible sources of bias in the setup and their countermeasures. In section \ref{extraction} we calculate the min-entropy for randomness extraction and discuss the extractor. Section \ref{results} contains results from the stringent statistical test suits applied on the random bit sequence for multiple runs. We conclude our work in section \ref{conclusion}.

\section{Optical source}\label{Optical source}
The optical field is described at quantum level in terms of photons. Fock states, $\left|n\right\rangle$ are states that contain $n$ photons with same frequency, polarization and temporal profile. Coherent states are represented as superposition of Fock states by:

\begin{equation}
\left|\alpha\right\rangle =e^{\frac{-\left|\alpha\right|^{2}}{2}} \sum_{n=0}^{\infty}\frac{\alpha^{n}}{\sqrt{n!}} \left|n\right\rangle
\label{coherent_state}
\end{equation}

where, $\alpha$ is a complex number. The amplitude $\left|\alpha\right|^{2}$ is the 
mean photon number of the state. A coherent state follow Poisson distribution, with high attenuation they can be approximated as single photon Fock states. 

\section{Source of randomness}\label{Source of randomness}

A Poisson process is used to model arrival times, which can be defined
as a sequence of random variable (rv) $0<S_{1}<S_{2}..S_{n}$ where $S_{i}<S_{i+1}$. Any arrival process $\left(S_{n}\right)$ can be
specified by inter-arrival times $\left\{ X_{i};i\geq1\right\} $ or
counting times $\left\{ N(t);t>0\right\} $ where,  $\left\{ S_{n}\leq t\right\} =\left\{ N(t)\geq n\right\} $.
$S_{n}$ is the sum of $n$ i.i.d rv each with exponential time distribution
$\lambda e^{- \lambda t}$. Hence, the density function of arrival
time at $n^{\textrm{th}}$ epoch \cite{gallager2013stochastic} is given   by

\begin{equation}
f_{S_{n}}(t)=\frac{\lambda^{n}t^{n-1}e^{- \lambda t}}{(n-1)!}.
\label{gamma_dist}
\end{equation}

This is a gamma distribution with $n$ and $\lambda$ parameter. We have considered arrival time of photon relative to an external reference clock. The exponential distribution of inter-arrival is flattened by choosing an external reference clock with the period (T), as proposed in \cite{Nie2014}. The number of photons arriving within a period T follows Poisson distribution. Thus, the probability of $n$ photons arriving in time T is given by:

\begin{equation}
p\{N(T)=n\}=\frac{(\lambda T)^{n}e^{-\lambda T}}{n!}.
\label{eqn: pois_dist}
\end{equation}

We further divide the T time period into $N_{b}$ small time segments called time bin $\tau_{1},\tau_{2},\tau_{3}....\tau_{N_{b}}$ where, $N_{b}$ is the total number of divisions and $\frac{T}{N_{b}}$ is the duration of each time bin. Precisely, $\left(i-1\right)t_{\textrm{bin}}\leq\tau_{i}\leq it_{\textrm{bin}}$ where $1\leq i\leq N_{b}$ and $t_{\textrm{bin}}=\frac{T}{N_{b}}$. Given a photon detection in $T$, we consider the arrival time of $n^{th}$ photon as $S_{n}=\tau$. With the detector dead-time greater then T, only single detection is possible, hence, $N(T)=1$. The probability distribution function (PDF) of $\tau_{i}$ conditioned on single detection at T is given by

\begin{equation}
\begin{split}
P(\tau_{i} | N(T) = 1) &= \frac{P( N(T- (i-1) t_{bin}) = 1, N (T - i t_{bin})=1)} {P(N(T) = 0)} \\
& = \frac{t_{bin}}{T}.
\end{split}
\end{equation}

Therefore, given a photon detected in $T$, the photon appears in each bin with equal probability, hence, desired uniform distribution is achieved.

\section{System architecture}\label{system_architechture}
\begin{figure}
\begin{center}
\includegraphics[scale=0.5]{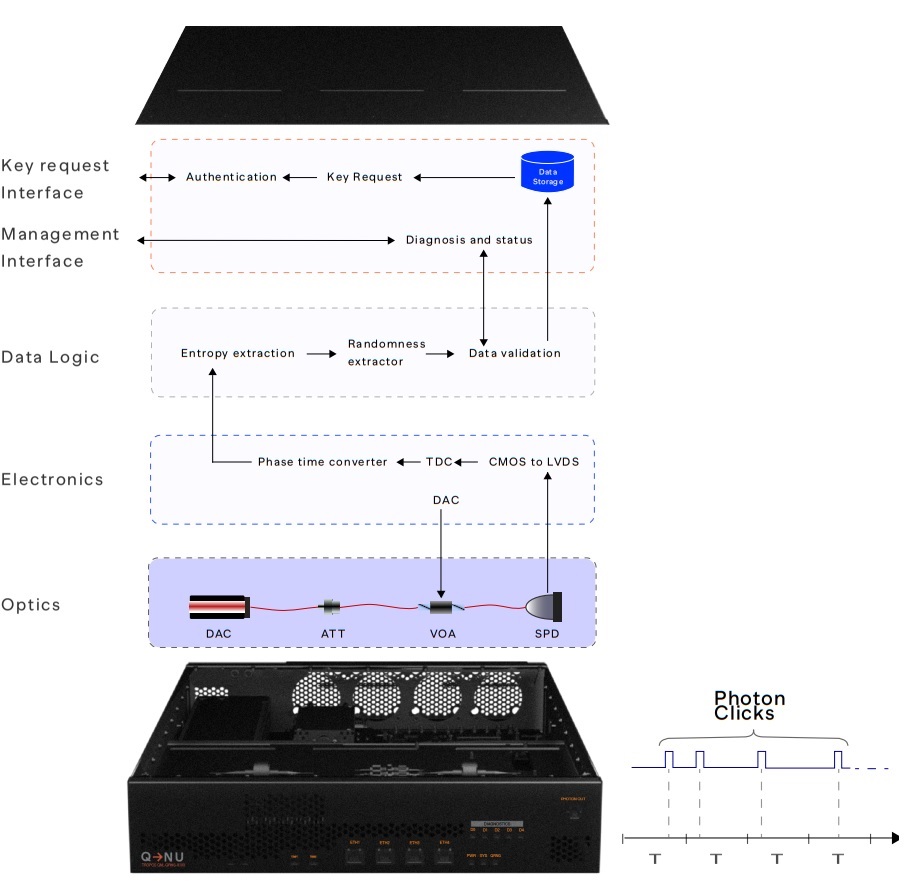}
\caption{Schematic for system architecture.}
\label{fig:System-architechture}
\end{center}
\end{figure}

In Fig. \ref{fig:System-architechture} we have presented the  self-contained QRNG architecture. We have divided it into 5 sections: (a) optics section consists of active and passive optical components, (b) electronics section  consists of high speed electronics comprising of  Field programmable gate arrays (FPGA) supporting driving electronics, (c) Data logic section consists of algorithms generating random number generation, binning, extraction and  data validation. In (d) we finally store the data and  perform key management and key request applications and in (e) we perform up-gradation of the system. In the optical path, we have used an SFP (small form-factor pluggable module) as a light source with $850$ nm wavelength operating in continuous mode. The optical source \cite{SFP} is heavily attenuated with an arrangement of fixed and variable attenuator providing flexibility to fine tune the optical amplitude with the resolution of 1 dbm. The arrangement enables proper calibration and tuning for  desired mean photon number $(<1)$.  We have used high-efficiency Silicon based single-photon detectors from Excelitas Technologies SPCM-AQRH \cite{spd}. The particular module under free-running mode is characterized by a detection efficiency of $38\%$, dead-time of $ T_d = 24$ ns, and dark count rate of $64$ cps. Once a photon is detected, SPD transmits an interrupt signal to the time-to-digital converter (TDC) chip. The TDC is configured to measure events with a precision of $1$ ps. The TDC output is given  to phase time extraction to identify and extract the fine time stamps. Once the time-stamps are recorded, we apply the randomness extraction algorithm where they are divided into bins and eventually converted into binary and post processing is applied. These algorithms are performed on  FPGA. In  validation section, entropy tests and optional (standard test suites) are performed which  determines the pass or fail criterion for generated sequence. If it passes all tests then we store the data. We can retrieve the status and diagnosis report  when required. Raw data or conditioned data can be procured from the system and evaluated separately. Finally, we have independent  ports for  key request interface and key  management interface. A separate track is considered for local up-gradation. The upgrade travels through authentication  of device, key validation, upgrade server. Thereafter, the system is upgraded.

\section{Scheme implementation}\label{implementation}

In the present QRNG, the reference clock period is set to 16.384 ns. The optical power  is set such that $\eta \lambda T = 0.7$, where $\eta$ is the detector efficiency. The fixed clock period, T, is then divided into $N_b (= 2^n)$ equal time bins $\tau_i$ as described in section \ref{Source of randomness}. Measuring the arrival time via an external clock is mathematically similar to performing a modulo with T. Thus, to achieve high precision and flexibility, a modulo operation of photon arrival time with time period T is performed. 

In section \ref{Source of randomness}, it is proven that the photons arriving in these short time bin $\tau_i$ have a uniform probability distribution of $1/N_b$. Thus, the uniform distribution of the random numbers generated via this scheme is theoretically guaranteed without any post-processing. For single photon detection in period T, it is essential that $T \leq T_d$. Also, the precision and jitter of the time measurements determine the upper bound on the number of time bins, $N_b$. 
To find an optimal value for T and $N_b$, we analysed different configurations on the basis of entropy per bit and coefficient of variance (CoV) in uniform distribution of number. The results are tabulated in Table \ref{analysis}. We have found that with $T = 16.384 $ ns and $N_b = 256$ least bias is observed. 

\begin{table}
\centering
\caption{Analysis for different configuration of T and $N_b$.  \label{analysis}}
\begin{tabular}{c c c c c }
\toprule
\textbf{S. No.} & \pmb{$N_b$} & \pmb{$t_{bin}$ (ps)} & \textbf{T (ns)} &  \textbf{CoV} \\
\midrule
1 &	256	& 78.125 & 20 & 0.018093926 \\
2 &	256	& 78 & 19.968 & 0.009843709 \\
3 &	512	& 39 & 19.968 & 0.012292352 \\
4 &	1024 & 19 & 19.456 & 0.022475629 \\
 5 & 256	& 64 & 16.384 & 0.009436647 \\
6 &	512	& 32 & 16.384 & 0.011936302 \\
7 &	1024 & 16 & 16.384 & 0.019921429 \\
\bottomrule
\end{tabular}
\end{table}

\section{Source of Bias} \label{bias}

A deeper understanding and further enhancements of present QRNG demands identifying and eliminating various biases and imperfections in the system. We have listed some of the imperfections in implementation affecting the randomness and also their countermeasures here:
\begin{enumerate}
\item A stable light source in optical QRNG corresponds to a reliable entropy source. Variations in optical power alter the photon statistic arriving at the detector, resulting in a deviation from the assumed scheme. To avoid any fluctuations optical source is temperature controlled and is supplied with high fidelity injection current. 
\item An untold assumption for a time of arrival QRNG is using a single-photon source. Due to practicality, an attenuated coherent source is used, creating a non-zero probability of multi-photon. However, we have considered the mean photon number less than unity, thus, considerably reducing the chances of multi-photon.
\item A practical single-photon detector has a limited detection efficiency ($\eta$) of less than $100\%$. For theoretical calculations, $\eta$ is considered a part of attenuation, while detectors are assumed to be $100\%$ efficient. Equivalently, the mean photon number equals $\lambda \eta $.
\item Non-zero dark count probability is yet another limitation/device-imperfection in SPDs. The dark count rate is the average rate of registered counts without incident photons. The Silicon based SPD has a  dark count rate of 64 clicks per second. For a better implementation of the scheme or uniform distribution of random digits, it will be appropriate to opt for lower dark count rate.
\item Avalanche-based SPD has a non-zero dead-time after each photon click where, the detector goes blind, limiting the maximum count rate. Wayne \textit{et al.} \cite{Wayne2009} showed that as the count rate approaches the inverse of the dead time, the majority of the photons will fall into the lower-valued time bins, reducing the entropy per click. The entropy rate peaks near half the maximum count rate of the detector. 
\item Timing jitter is defined as the uncertainty in registering the timing of the photon events. The two sources of time jitter in the current setup are SPD and TDC. Choosing a bin size higher than the maximum jitter would eliminate bias due to time-uncertainty of detector. In this work, the randomness is also contributed by the timing jitter, however,  this is not a technology limitation and can be circumvented by upgrading the detector.
\end{enumerate}

\section{Randomness estimation and extraction}\label{extraction}
\begin{figure}
\centering
\includegraphics[scale=0.3]{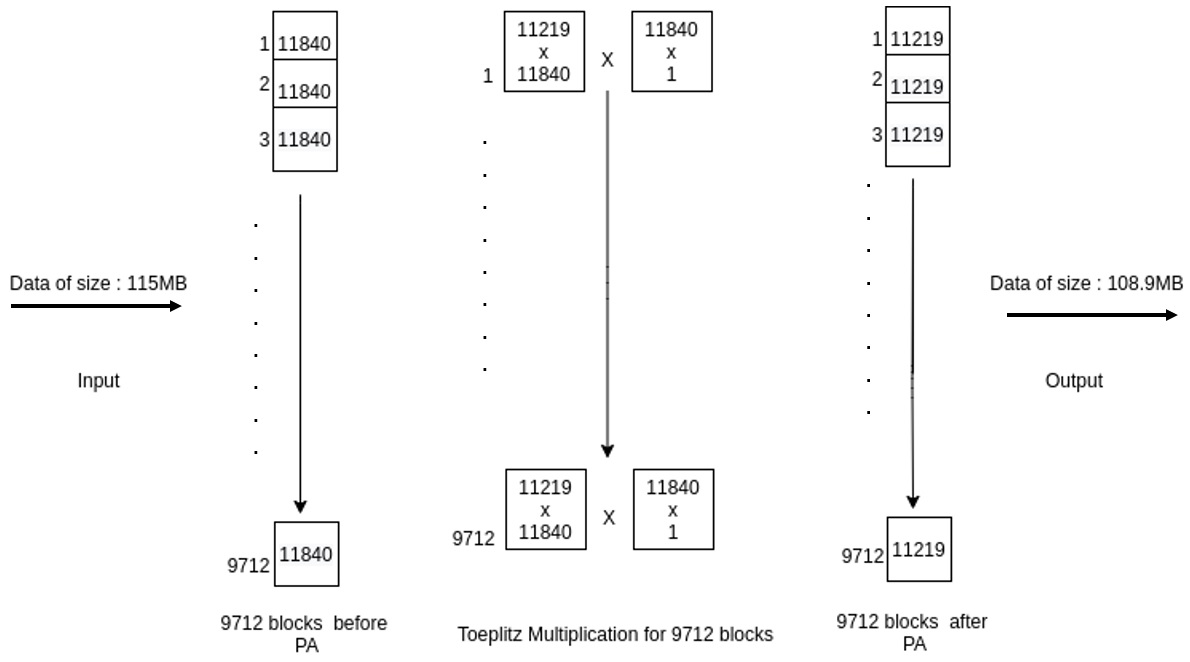}
\caption{Randomness extraction methodology}
\label{fig:QRNG_Implementation}
\end{figure}

Evaluating the  min-entropy of the raw data helps in quantifying the randomness of the data. The min entropy is defined as:
\begin{equation}
H_{\textrm{min}}(X)=-\textrm{log}_{2}\,(\textrm{max}_{x\in{0,1}^{N}}P_{r}[X=x])
\end{equation}

which can be utilized to estimate the extraction ratio between raw
random bits and the final random bits. After the  min-entropy evaluation,
a Toeplitz hashing is performed  to filter the raw data. Toeplitz hashing is a randomness extractor function that  produces almost uniform string as the
output. The extractor makes use of an additional independent random
seed as input called seeded extractors. This function is given by,
\begin{equation}
\textrm{EXT}:\left\{ 0,1\right\} ^{n}\times\left\{ 0,1\right\} ^{d}\rightarrow\left\{ 0,1\right\} ^{m}
\end{equation}

which takes $n$ bits from the raw sequence as input and a uniform
random seed that is of $d$ bits to produce an output of $m$ bits.
We call a $(k,\epsilon)$-extractor to a function, for any input $k$ source (a raw sequence of at least, min-entropy $k$), to produce final
sequence that is \ensuremath{\epsilon}-close to uniform. Given a binary Toeplitz matrix of size $m\times n$, $m$ random
bits are obtained as result of multiplying the matrix and $n$ raw
bits. The output sequence of length $m$ depends on the following
parameters:
\begin{itemize}
\item $n$- size of raw data
\item $k$- min-entropy of raw data
\item $\epsilon$-security parameter
\end{itemize}
By substituting the above parameters in the equation of leftover hash
lemma we could determine the length of $m$ random bits as
\begin{equation}
m=n\,H_{\textrm{min}}(X)-2\,\textrm{log}_{2}(1/\epsilon)
\end{equation}

The Toeplitz matrix can be constructed by a sequence of $m+n-1$ random
bits due to its characteristic that all the elements of each descending
diagonal from left to right are the same. In the usual Toeplitz matrix,
its security is based on the randomness of the elements in the first
column and the first row. In LFSR-based Toeplitz, we only initialize
the first column of the Toeplitz matrix with a seed generated from
PRNG/in built QRNG, and the LFSR outputs a sequence that acts as the first row of
the Toeplitz. The LFSR considers the initialized column and the chosen
irreducible polynomial as input to produce the sequence.

In Figure \ref{fig:QRNG_Implementation}, we have presented an implementation of extraction, where
the data of size $115$ Mb is divided into $9712$ blocks. Each block
is of size $11840$. The Toeplitz matrix of size $11219 \times 11840$
has been constructed. Further, the bit(s) multiplication happens for
each of the 9712 blocks with the Toeplitz matrix. Finally, the output
of each multiplication process is concatenated to obtain the final
output i.e., $11219$ bits from each block which results in $11219 \times 9712$
= $108.9$ Mb. The extraction ratio  comes out to be $0.94$ and
the the extraction efficiency is $0.89$. The corresponding extraction
ratio is smaller than the min-entropy of the raw data. The min-entropy
of the raw data ranges from $0.95-0.99$. Therefore, the randomness
of the extracted quantum random number can be guaranteed. Considering
a single block of Toeplitz implementation, we have chosen $m=11219,n=11840$
and the worst case min-entropy of $0.95$ is considered by substituting
the values in the equation of leftover hash lemma one can calculate
the information theoretic security bound $(\epsilon)$ which is given by $2^{-15}$.

\section{Randomness  evaluation} \label{results}

Assessing randomness of  RNG is a well studied subject and numerous statistical tools are present to categorically determine the quality of randomness. We have adopted a strategy of extensive testing of our QRNG via all the available stringent test suites including: NIST SP 800-22, Dieharder, ENT and Test U-01. Apart from this we also perform statistical test like: entropy estimation, auto-correlation and probability distribution. The results of individual tests and their implication is present below.

\subsection{Auto Correlation}
\begin{figure}
\centering
\includegraphics[scale=0.18]{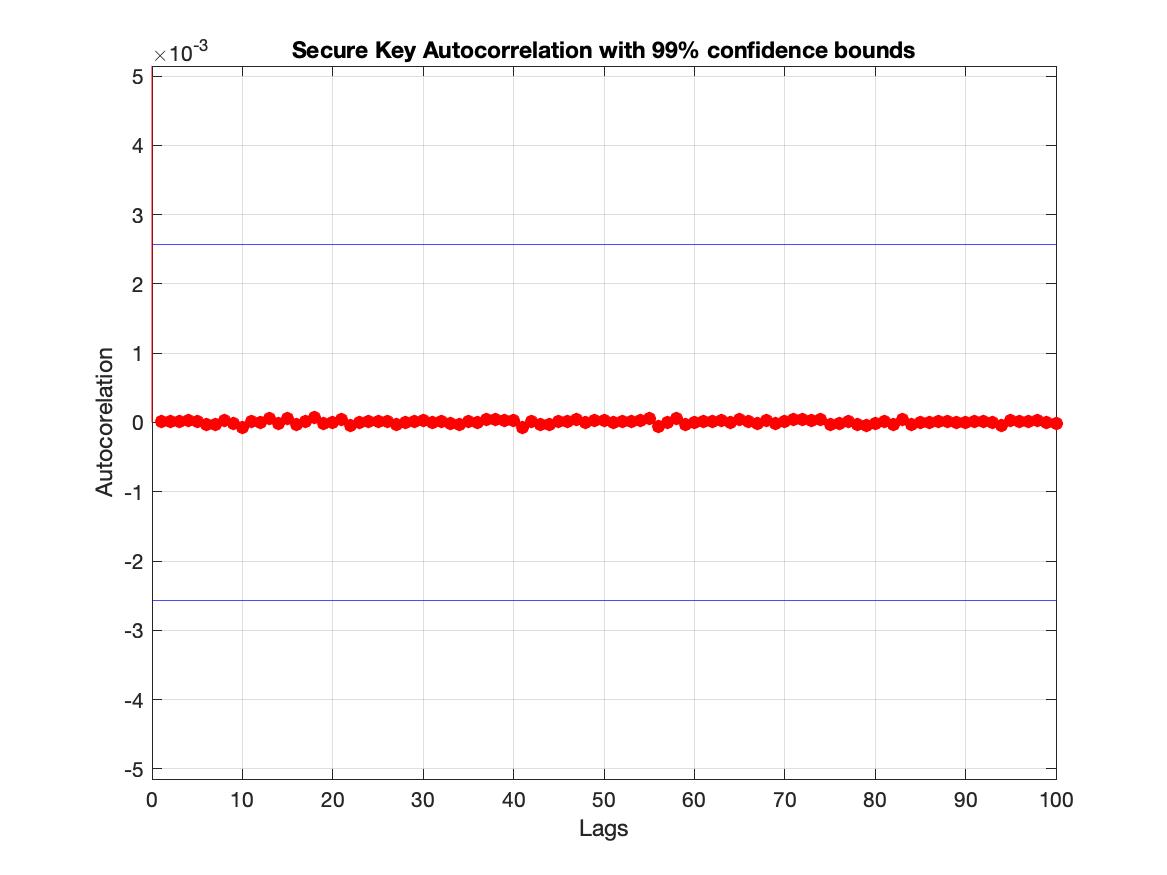}
\caption{Auto correlation plot with Lag = 100.}
\label{autocorr}
\end{figure}

Auto correlation is a widely used technique in signal processing to find repeating patterns or periodic signal. It finds the correlation of a signal with a delayed copy of itself as a function of delay. A similar analysis is also useful to check the correlation between random number or to identify patterns. We plot (Fig. \ref{autocorr}) the correlation of numbers with a lag (delay) of upto 100 numbers along with 99\% bounds. 

\subsection{Probability distribution}
\begin{figure}
\centering
\includegraphics[scale=0.18]{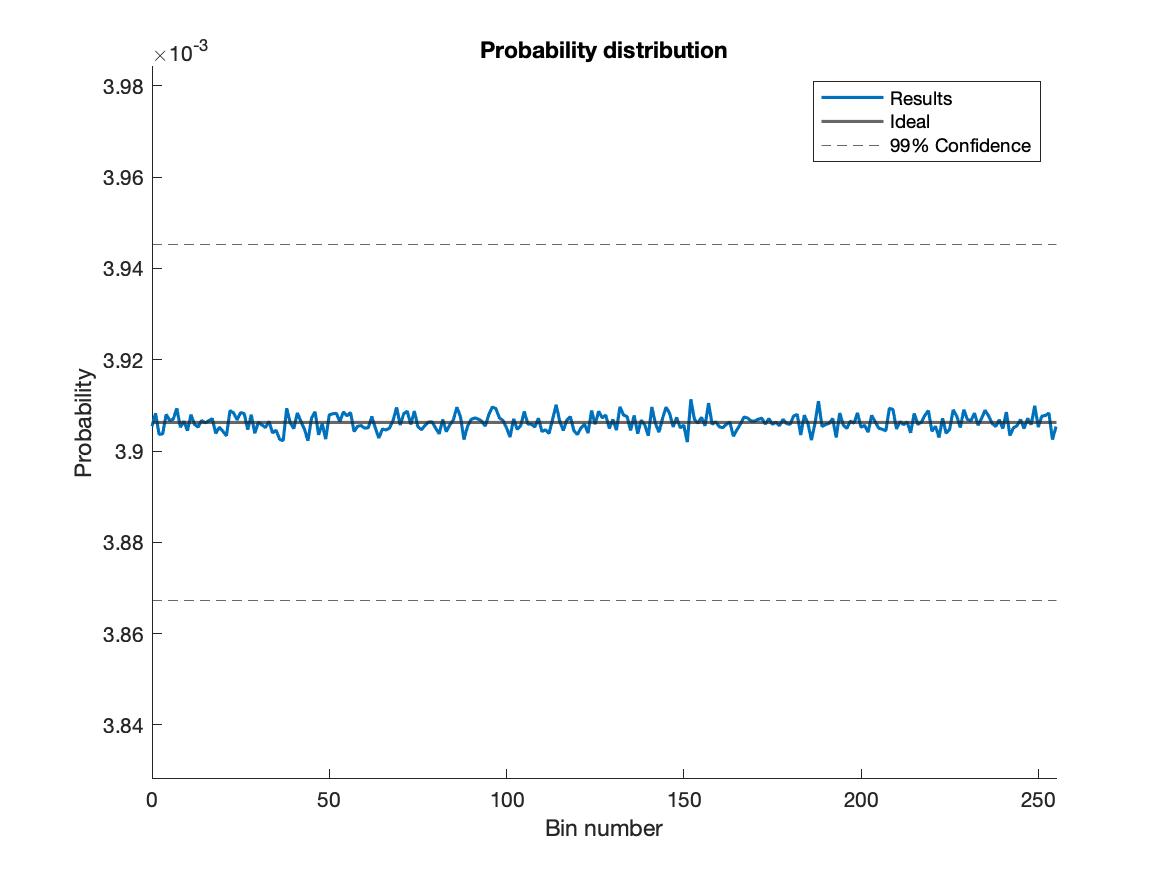}
\caption{Distribution of probability of time bins from 0-255.}
\label{hist}
\end{figure}

Apart from this, we also plot $p_i$ for different bin number as shown in Fig. \ref{hist} and calculate the coefficient of variance for $p_i$. This exercise help us visualise the uniform distribution of the probability and quantify the variance due to imperfections

\subsection{ENT}
ENT \cite{ENT} is a test-suite primarily designed to evaluate randomness of PRNG. First test  is entropy which calculates the number of bits of information contained in the byte(bit).  TRNGs are expected to yield greater than 7.9.(1) bits of information per byte(bit).  This decides the compression of the file which should not be zero. The Chi square test is the most stringent test to determine the uniformity of data. This is an important metric to indicate bias in system. The arithmetic mean performs just what the name suggest. We consider it to be 0.5 at bit level  and at the byte level it should be close to 127.5 for the data to be uniform.  This test is followed by Monte Carlo method.  The serial correlation coefficient,  indicates dependence of the present value with the previous value and it is expected to be as close to 0 and less than 0.005 for TRNG. In Table \ref{ent}, we have compared  results with expected  ideal result.

\begin{table}
\caption{ENT test results}
\begin{tabular}{p{0.45\textwidth} p{0.225\textwidth} p{0.225\textwidth}}
\toprule
\textbf{Variable} & \textbf{QRNG} & \textbf{Ideal} \\
\midrule
Entropy (per bit) & 1.0000 & 1.0000 \\
Chi-square distribution & 55.95\% & 10\% \textasciitilde 90\% \\
Arithmetic mean value & 0.4980 & 0.5000 \\
Monte Carlo value for $\pi$ & 3.141513243 & 3.1415926536 \\
Serial correlation coefficient & 0.0000065 & 0.000000 \\ \bottomrule
\end{tabular}
\label{ent}
\end{table}

\subsection{NIST Test Suite}
NIST’s STS (SP 800-22 suite)  \cite{nist2,nist1} comprises of 15 statistical tests formulated to assess the specific null hypothesis of the sequence being random. 
Analysis of results  is approached by  two methods namely,
\begin{enumerate}
    \item proportion of Sequences Passing a Test and 
    \item uniform distribution of P-values
\end{enumerate}
The NIST standard defines a range of allowable proportion using the confidence interval,
\begin{equation}
    p\pm3\sqrt{\frac{p(1-p)}{m}}
\end{equation}

where $p=1-\alpha$ and $m$ is the sample size. If the proportion pitch outside of this space, then there is an indication that the data is non-random. For a sample size of 1000, with the significance level $(\alpha)$ as 0.01, the range of acceptable proportion defined using the confidence interval is $ .99\pm0.0094392$ (i.e.,above 980). 

\begin{figure}
\centering
\begin{tabular}{cc}
\includegraphics[width = 0.6\textwidth]{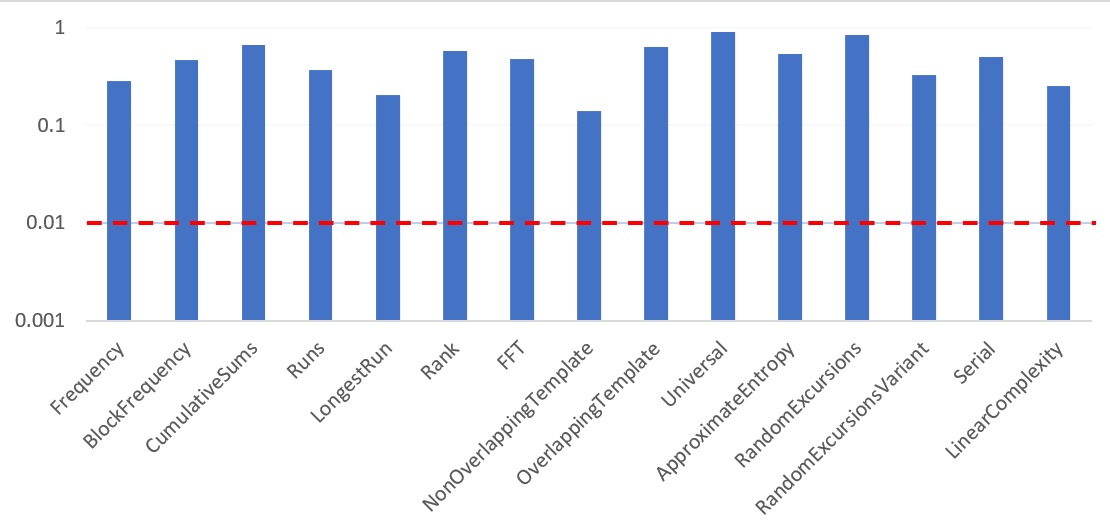}
\end{tabular}
\caption{P-value for NIST test results}
\label{nistpvalue}
\end{figure}

\begin{figure}
\centering
\begin{tabular}{cc}
\includegraphics[width = 0.6\textwidth]{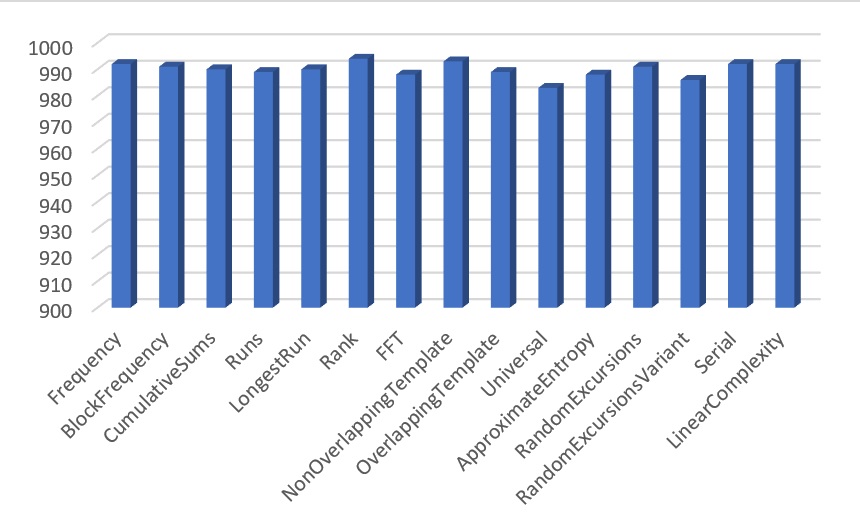}
\end{tabular}
\caption{Confidence plot for NIST test results}
\label{nistconfidence}
\end{figure}

Individual test calculates a statistic value using the input data. The test statistic is used to calculate a P-value that summarizes the strength of the evidence against the null hypothesis. A significance level  is chosen for the tests. If P-value is greater than of equal to significance value then the null hypothesis is accepted; i.e., the sequence appears to be random else, the sequence appears to be non-random. Results of the suite averaged over 100 files of 1 Gb each are presented in Fig. \ref{nistpvalue} and the confidence ratio is presented in  Fig. \ref{nistconfidence}.


\subsection{Diehard}
\begin{figure}
\centering
\begin{tabular}{cc}
\includegraphics[width = 0.6\textwidth]{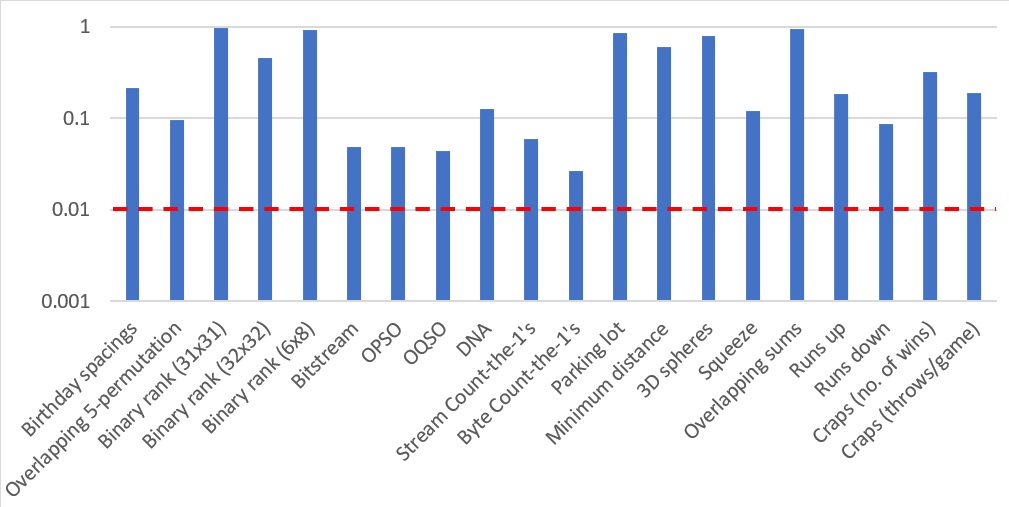}
\end{tabular}
\caption{P-vale Diehard results}
\label{diehard}
\end{figure}

We have evaluated  the generated bit stream with Diehard test suite .  One billion continuous bit streams were tested and in  Fig. \ref{diehard} we have presented the results.

\subsection{Dieharder}
\begin{figure}
\centering
\begin{tabular}{cc}
\includegraphics[width = 0.6\textwidth]{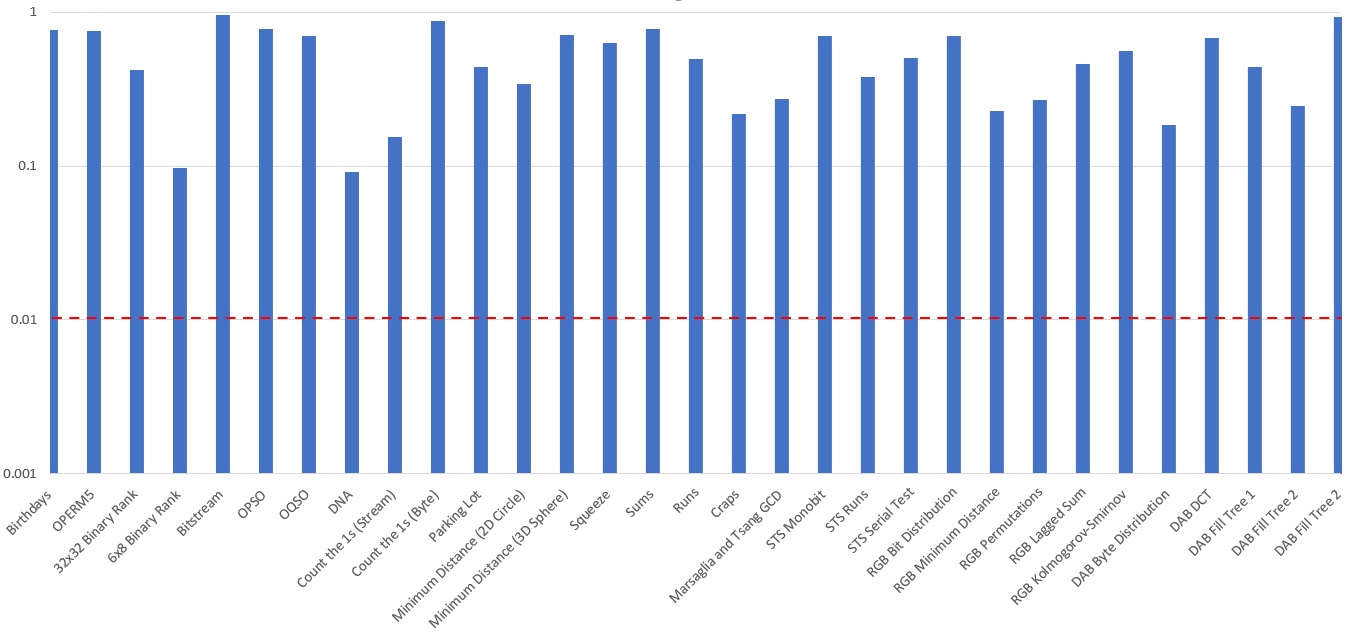}
\end{tabular}
\caption{P-vale Dieharder results}
\label{dieharder}
\end{figure}

Dieharder \cite{dieharder} is known as the gold standard for evaluating RNG. The Dieharder battery of tests is a collection of 31 individual tests. Similar to NIST test, it returns p-value for the hypothesis testing weather the given sequence is random or not. Additionally, Dieharder convert a set of P-values into a single P-value by comparing their distribution to the expected one, using a Kolmogorov-Smirnov test against the expected uniform distribution of P. Dieharder classifies the test results as passed, weak or fail. We have summarised the results on our QRNG for over 100 Gb of data in Fig. \ref{dieharder}.

\subsection{Test U-01}
The TestU-01 suite \cite{testu01} is one of the most stringent collection of tests to assess RNG statistical properties. TestU-01 Small Crush has many tests common  with Dieharder, SP 800-22 and  FIPS 140-2, and tests recommended under AIS-31 (via appended recommendations documents). The results are presented in Table. \ref{TestU01}.

\begin{table}
\centering
\caption{Test U-01 test results}
\begin{tabular}{p{0.45\textwidth} p{0.12\textwidth}   }
\toprule
\textbf{Test Name} & \textbf{Result}  \\
\midrule
Rabbit Battery Test & All Passed \\ 
Alphabit Battery Test & All Passed \\ 
FIPS 140-2  & All Passed \\ 
Small Crush Battery Tests & All Passed \\ \bottomrule
\end{tabular}
\label{TestU01}
\end{table}

\begin{table}
\centering
\caption{Bench-marking the results with existing works.}
\begin{tabular}{p{0.25\textwidth} p{0.14\textwidth} p{0.12\textwidth} p{0.14\textwidth} }
\toprule
 &\textbf{Banerjee\textit{ et al.}} &  \textbf{Nie \textit{et al.}} \cite{Nie2014}  & \textbf{Yan \textit{et al.}} \cite{yan}  \\
\midrule
Raw Data (Mbps) & 115  & 109 & 128\\ 
Bit generation efficiency & 8 & 8 & 8\\ 
Min-entropy & .95  & .88 & \\ 
Conditioned data & 109 & 96 &\\ 
Test Suites & NIST,ENT,  & NIST, ENT & NIST\\
 &  Diehard,TU-01 &  & \\
 &   Dieharder &  & \\ \bottomrule
\end{tabular}
\label{benchmark}
\end{table}

\section{Conclusion} \label{conclusion}
Summarizing the experimental results we  report generation of high entropy and consistently uniform random numbers from quantum entropy based on  arrival time of photons relative to an external reference time period. In Table \ref{benchmark}, we have compared the results with existing work and it is found it be consistent with the expectations. We have investigated the randomness of the generator before and after conditioning and examined its sources of bias. We have briefly discussed   Toeplitz hashing implementation in FPGA. We report the  statistical distance between the extracted random sequence and the
uniform sequence is bounded by $(\epsilon)$ which is given by $2^{-15}$. The extraction
efficiency is $0.89$.  We have validated the results extensively against stringent randomness test suites namely NIST, ENT, Diehard, Dieharder, TU-01 and the results are consistently encouraging.

{}


\begin{thebibliography}{}
\bibitem{HerreroCollantes}M. Herrero-Collantes and J. C. Garcia-Escartin, Quantum random number generators, Rev. Mod. Phys. 89, 015004 (2017).

\bibitem{ishida1956} M. Ishida,  Random number generator, Annals of Institute of Statistical Mathematics, 8 (119--126) 1956.


\bibitem{Wayne2009} M. A. Wayne, E. R. Jeffrey, G. M. Akselrod and P. G. Kwiat, Photon arrival time quantum random number generation, J. Mod. Opt. 56:4, 516--522 (2009).
\bibitem{Wahl2011} M. Wahl, M. Leifgen, M. Berlin, T. Röhlicke, H.-J. Rahn and O. Benson, An ultrafast quantum random number generator with provably bounded output bias based on photon arrival time measurements, Appl. Phys. Lett. 98, 171105 (2011).
\bibitem{Li2013} S. Li, L. Wang, L.-An Wu, H.-Q. Ma, and G.-J. Zhai, True random number generator based on discretized encoding of the time interval between photons, J. Opt. Soc. Am. A 30, 124 (2013).
\bibitem{Wayne2010} M. A. Wayne and P. G. Kwait, Low-bias high-speed quantum random number generator via shaped optical pulses, Opt. Express 18, 9351 (2010).
\bibitem{Ma2005} Hai-Qiang Ma, Yuejian Xie, and Ling-An Wu, Random number generation based on the time of arrival of single photons, Appl. Opt. 44, 7760-7763 (2005)
\bibitem{Nie2014}Y. Nie, H. Zhang, Z. Zhang, J. Wang, X. Ma, J. Zhang and J. Pan, Practical and fast quantum random number generation based on photon arrival time relative to external reference, Appl. Phys. Lett. 104, 051110 (2014).
\bibitem{gallager2013stochastic} R. G. Gallager, Stochastic processes: theory for applications,Cambridge University Press, 2013.


\bibitem{spd} Excelitas Technologies’ Single Photon Counting Module, https://www.excelitas.com/product/spcm-aqrh

\bibitem{SFP} Finisar’s FTLX8574D3BCL 10Gb/s SFP module, https://ii-vi.com/product/10gbase-sr-sw-400m-multimode-datacom-sfp-optical-transceiver/

\bibitem{ENT}J. Walker, ENT: A Pseudorandom Number Sequence Test Program, http://www.fourmilab.ch/random.


\bibitem{nist2}  NIST: Random Number Generation and Testing, https://csrc.nist.gov/Projects/Random-Bit-Generation/Documentation-and-Software.

\bibitem{nist1} A. Rukhin, J. Soto, J. Nechvatal, M. Smid, E. Barker, S. Leigh, M. Levenson, M. Vangel, D. Banks, A. Heckert, J. Dray and S. Vo, NIST Special Publication 800-22 (NIST, 2008), http://csrc.nist.gov/rng/.
\bibitem{dieharder} G. Marsaglia, DIEHARD: a battery of tests of randomness, http://stat. fsu. edu/geo, 1996.
\bibitem{testu01} P. L'Ecuyer and R. Simard, TestU01: AC library for empirical testing of random number generators, ACM Transactions on Mathematical Software (TOMS), ACM New York, NY, USA 33(1-40) 2007.


\bibitem{yan} Q. Yan, B. Zhao, Z. Hua, Q.g Liao, and H. Yang, High-speed quantum-random number generation by continuous measurement of arrival time of photons, Review of Scientific Instruments 86, 073113 (2015)

\end{thebibliography}
\end{document}